\documentclass{emulateapj}

\newcommand{\G}{$G$}

\newcommand{\eg}{{e.g. \/}}
\newcommand{\ie}{{i.e. \/}}

\newcommand{\etal}{{et al. \/}}

\newcommand{\um}{$\mu$m}

\slugcomment{Publications of the Astronomical Society of the Pacific, in press}

\shorttitle{Independent Photo-$z$s with Euclid}
\shortauthors{Sorba \& Sawicki}

\begin{document}


%\title{How On-board Optical Wavelength Photometry can Improve Photometric Redshifts with Euclid}
\title{How Future Space-Based Weak Lensing Surveys might Obtain Photometric Redshifts Independently}


\author{R. Sorba and M. Sawicki}
\affil{ Department of Astronomy and Physics and Institute for Computational Astrophysics, Saint Mary's University, 923 Robie Street, Halifax,
Nova Scotia, B3H 3C3, Canada }
\email{rsorba@ap.smu.ca, sawicki@ap.smu.ca}


%\altaffiltext{1}{Visiting Astronomer, Cerro Tololo Inter-American Observatory.CTIO is operated by AURA, Inc.\ under contract to the National ScienceFoundation.}
%\altaffiltext{2}{Society of Fellows, Harvard University.}
%\altaffiltext{3}{present address: Center for Astrophysics, 60 Garden Street, Cambridge, MA 02138}
%\altaffiltext{4}{Visiting Programmer, Space Telescope Science Institute}
%\altaffiltext{5}{Patron, Alonso's Bar and Grill}


\begin{abstract}
We study how the addition of on-board optical photometric bands to future space-based weak lensing instruments could affect the photometric redshift estimation of galaxies, and hence improve estimations of the dark energy parameters through weak lensing. Basing our study on the current proposed Euclid configuration and using a mock catalog of galaxy observations, various on-board options are tested and compared with the use of ground-based observations from the Large Synoptic Survey Telescope (LSST) and Pan-STARRS. Comparisons are made through the use of the dark energy Figure of Merit, which provides a quantifiable measure of the change in the quality of the scientific results that can be obtained in each scenario. Effects of systematic offsets between LSST and Euclid photometric calibration are also studied. We find that adding two ($U$ and $G$) or even one ($U$) on-board optical band-passes to the space-based infrared instrument greatly improves its photometric redshift performance, bringing it close to the level that would be achieved by combining observations from both space-based and ground-based surveys while freeing the space mission from reliance on external datasets.  
\end{abstract}




%% in the electronic edition to the object data available at the data cente%%

\section{Introduction}
\label{intro}
Over the past few years, it has been shown that approximately 74\% of the energy density of the universe is in the form of dark energy (DE). Often represented as the cosmological constant $\Lambda$ in Einstein's theory of general relativity, DE causes the expansion of the Universe to accelerate (see Copeland \etal 2006 for a review). A great deal of effort is being put into constraining the DE equation of state parameters in order to better understand the phenomenon. One promising method of placing high accuracy constraints on the DE parameters is through weak lensing, which involves measuring the shape of numerous galaxies over a large area of the sky (\eg Blandford \etal 1991, Bartelmann \& Schneider 2001, Refregier 2003).

An important source of error in the weak lensing analysis comes from uncertainties in the photometric redshift (photo-$z$) estimation to each galaxy, which is required to form a three dimensional map. While they are less accurate than spectroscopic redshifts (spec-$z$s), the extremely large area (20 000 deg$^2$) planned for future dark energy surveys necessitates the use of photo-$z$s (see Hildebrandt \etal 2010 for a summary of current photometric redshift techniques and capabilities). The effects of photo-$z$ uncertainities on weak lensing were studied by Huturer \etal (2006), who placed stringent constraints on the degree of accuracy and precision required in order for the redshift estimations to be useful in tomography. Indeed, current goals state that the standard deviation of the photo-$z$s must be less than 0.05($1+z$) and that any bias in the photo-$z$s must be known to a degree of 0.002($1+z$). Ma \etal (2006) found that in order to satisfy these criteria, a large spectroscopic survey (on the order of $\sim10^5$ galaxies) must be carried out to properly calibrate the photometric redshifts. Several works since then have found that the amount of spectroscopy needed can be reduced by optimizing the spectroscopic survey to cover important redshift ranges (Ma \& Bernstein 2008, Sun \etal 2009, Berstein \& Huterer 2010), while Bordoloi \etal (2010) studied how the photo-$z$s themselves can be used for calibration.

Because of confusion between breaks, photometric redshifts are especially prone to catastrophic outliers, which can greatly impact the weak lensing analysis (Sun \etal 2009, Bernstein \& Huterer 2010). It was found that the number of catastrophic outliers can be greatly reduced if ground-based photometry (u,g,r,i,z,y) is complimented with near-infrared (NIR) photometry (Abdalla \etal 2008, Nishizawa \etal 2010). To this aim, both WFIRST and Euclid, two space-based instruments that will study weak lensing, will include a NIR channel to facilitate accurate photometric redshifts. However, in their current state, neither instrument can obtain accurate photo-$z$s by itself. Instead, they must rely on complimentary observations from ground-based instruments (such as the proposed Large Synoptic Survey Telescope, LSST, or the Panoramic Survey Telescope and Rapid Response System, Pan-STARRS). Various complications may arise from trying to combine such a large amount of data from two or more observatories, and it may be beneficial if a space-based instrument could produce its own optical observations.

In this work, we study how future space-based weak lensing missions may benefit from the addition of on-board optical photometry. We do this in the context of the current Euclid design in order to ground our results in reality, but the conclusions could apply just as well to WFIRST or any other future weak lensing instrument. Euclid is comprised of both a visual (VIS) channel and a near infrared (NIR) channel. The proposed VIS channel is made of a very broad filter ($RIZ$) which covers the wavelength range 0.55-0.92\um\ and will be used primarily to measure galaxy shapes for weak lensing. The NIR photometric channel includes three band-passes ($Y, J$, and $H$) and spans wavelengths from 1.0 to 1.6\um.  

While one could design a highly-optimized, multi-element on-board filter system, our aim, instead, is to explore simple scenarios that result in relatively small perturbations to the current Euclid instrument and mission design. To this end we explore the impact of adding two on-board optical band-passes ($U$ and \G) to Euclid and compare them to the results one would obtain by augmenting Euclid observations with ground-based optical observations. In Section 2, we detail the method with which we simulate a catalog of observed galaxies and generate photometric redshifts for each of the galaxies. In Section 3 we present the photometric redshift distribution of various filter combinations and compare these with the expected results of combining observations from Euclid with either LSST or Pan-STARRS using the dark energy Figure of Merit (FoM). Finally, we discuss the effects of any systematic offsets between the Euclid and ground-based observations on the weak lensing analysis.

%%%%%%%%%%%%%%%%%%%%%%%%%%%%%%%%%%%%%%%%%%%%%%%%%%%%%%%%
%                   SECTION 2
%%%%%%%%%%%%%%%%%%%%%%%%%%%%%%%%%%%%%%%%%%%%%%%%%%%%%%%%
\section[Method]{Method}
\label{method}

\subsection{Mock Catalog}
\label{method:catalog}

In order to study the effect of space-based optical observations, we first needed a catalog of galaxies to be observed. Ideally, this catalog would have realistic redshift, color, and luminosity distributions in order to accurately model the galaxy population and its photometric redshift distribution. To this aim, we chose to use the COSMOS Mock Catalog (CMC; Jouvel \etal 2009). This catalog draws upon observations from the COSMOS Deep Field (Capak \etal 2007) and the photometric redshift catalog of those galaxies (Ilbert \etal 2009). By combining the best fitting redshift and extinction with observable properties such as galaxy type and half-light radius (Leauthaud \etal 2007), the CMC is by construction representative of a real galaxy survey. 

The CMC best-fitting spectral energy distribution (SED) templates were generated similarly to Ilbert \etal (2009) which used template libraries of both Polletta \etal (2007) and Bruzual \& Charlot (2003). The Polletta \etal (2007) templates include SEDs of elliptical and spiral galaxies, whereas the Bruzual \& Charlot (2003) templates model starburst galaxies with ages ranging from 3 to 0.03 Gyr. Additional extinction was applied to the templates using the Calzetti \etal (2000) extinction law with $E(B-V)$ values of 0, 0.05, 0.1, 0.15, 0.2, 0.35, 0.3, 0.4, and 0.5.     

In summary, the CMC provides best-fitting spectra and observed properties of 538 000 galaxies from the COSMOS-ACS catalog and covers an effective area of 1.24 deg$^2$. It has a maximum redshift of $z=3.64$ and a median redshift of 0.96. The COSMOS mock catalog is limited
by the completeness of the COSMOS imaging
($i^+_{AB} \sim 26.2$ for 5$\sigma$ detection, Capak \etal 2007).

\subsection{Simulating Observations}
\label{method:observations}

The proposed observational strategy for Euclid involves taking 4 dithered exposures of each field, with the total integration time of each dither being approximately 700s (Duvet 2010). Euclid is designed such that the VIS and NIR channels can observe simultaneously. In order to limit the number of moving parts during the VIS integration, the NIR spectrometer and the VIS channel will observe first (with an integration time of $\sim$ 540s) after which the NIR photometry bands will observe sequentially (integration times given in Table \ref{tab:filters}). All of our simulations for Euclid observations assume this observational strategy.

We studied photo-$z$ performance with several Euclid $U$+$G$ channel configurations as outlined in a recent CSA-sponsored study (Rowlands \etal 2011). We first assumed a best-case scenario where two dichroics split the VIS channel light such that the $RIZ$, $G$, and $U$ filters can all observe simultaneously (\ie three separate detectors). However, we also tested other scenarios: one where we removed one detector and dichroic and put the $U$ and $G$ filters on a filter exchanger so that they must share integration time, and another where all the filters are on an exchanger and all feed the same single detector.  Finally, we also considered scenarios in which only one additional on-board filter --- either $U$ or $G$ --- is used. 

Assuming that the best-fitting SED from a galaxy in the CMC is the ``true'' signal from the galaxy, we simulated the observed magnitude of each galaxy in each of the Euclid filters (properties shown in Table \ref{tab:filters}) by adding a random noise component to the ``true'' signal measured through each filter.  

Our method of estimating noise follows closely that described in the appendix of Jouvel \etal (2010), but a brief outline is given here. First, we calculated the expected signal to noise ratio ($S/N$). For space-based observations, the $S/N$ can be found by
\begin{equation}
\frac{S}{N} =   {e_{\rm signal}\over{\sqrt{e_{\rm signal} + e_{\rm sky} + N_{\rm pix}N_{\rm exp}{e_{\rm RON}}^{2} + N_{\rm pix}N_{\rm exp}t_{\rm obs}e_{\rm dark}}}}.\
\end{equation}

Here, $e_{\rm signal}$ is the number of electrons produced in the device by the galaxy flux. The noise contributions in the denominator include Poissonian noise from the source as well as the background zodiacal light ($e_{\rm sky}$), the dark current caused by the thermal radiation of the instrument ($e_{\rm dark}$), and also the read-out noise of the detector ($e_{\rm RON}$) which follows a Gaussian statistic. $N_{\rm pix}$ is defined to be the number of pixels contained within a circular area of 1.4 times the observed full width at half maximum of the galaxy, and $N_{\rm exp}$ is the number of exposures (4 as planned in the Euclid survey) and $t_{\rm obs}$ is the exposure time. For the VIS channel, we use $e_{\rm dark}$=0.03 electrons per second, $e_{\rm RON}$ = 6 electrons, and assume 0.1 arcsecond pixels, whereas for the NIR we assume $e_{\rm dark}$=0.05 electrons per second, $e_{\rm RON}$ = 5 electrons, and 0.3 arcsecond pixels. The on-board optical $U$ and $G$ channel is assumed to have the same pixel scale as the NIR channel (0.3 arcsecond pixels), but the RON and dark current of the VIS channel.

\begin{figure}
\begin{center}
\includegraphics[width=8cm]{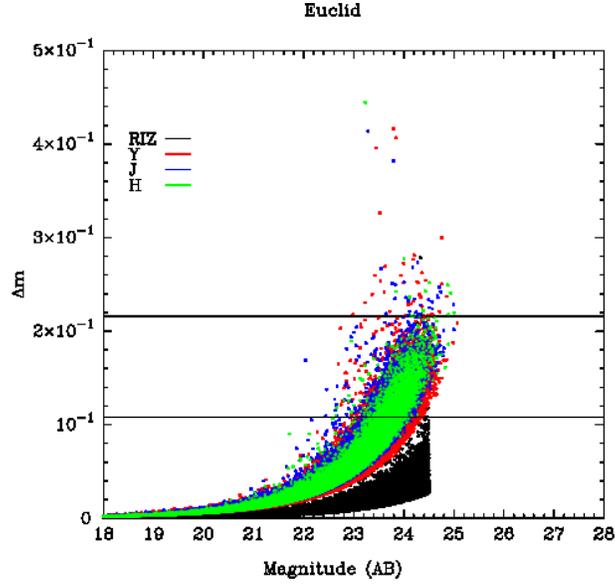}
\end{center}
\caption{\label{fig:error}Uncertainty verses ``observed'' magnitude for objects with $RIZ$ magnitude less than 24.5. Horizontal lines mark the 10-$\sigma$ and 5-$\sigma$ uncertainties.}
\end{figure}

Once the theoretical $S/N$ is determined, we simply added an error term to the true magnitude of the galaxy which is drawn from a Gaussian distribution of mean $\mu=0$ and standard deviation $\sigma={{2.5\over{\ln{(10)}}}{1\over{S/N}}}$. In this way, realistic observational uncertainties are included in our final simulated observations, as is demonstrated in Figure \ref{fig:error}. By comparing Figure \ref{fig:error} with results from other radiometric performance simulations of Euclid (the Euclid Reference Payload Concept Document (Duvet 2010) predicts a $S/N$ of 14.3 for the $RIZ$ filter at a magnitude of 24.5, and 7.1 for the IR filters at magnitude 24), we are confident that our noise generation procedure produces reasonable results.

For ground-based LSST observations our error simulations followed a different approach. In accordance with Ivezic \etal (2008), the expected photometric error for a single observation of a galaxy is given by 
\begin{equation}
\label{eqn:LSSTerr}
{\sigma_{LSST}}^2 = {\sigma_{sys}^2} + {\sigma_{rand}^2}
\end{equation}
where $\sigma_{sys} = 0.003$ and $\sigma_{rand}$ is given by
\begin{equation}
{\sigma_{rand}}^2 = (0.04-\gamma)x + \gamma{x^2}
\end{equation}
with $x = 10^{0.4(m-m_5)}$. Here $m_5$ is the 5$\sigma$ depth for point sources in a given band and $\gamma$ is based on factors such as the sky brightness and readout noise. Values for $m_5$ and $\gamma$ can be found in Table \ref{tab:LSST_params}. To account for repeat observations, $\sigma_{rand}$ is divided by 10 to give the error after 100 observations. In a similar fashion to the space based observations above, an error term was added to each of the model magnitudes, which is drawn from a Gaussian distribution but now with $\sigma={\sigma_{LSST}\over10}$. Gaussian errors for Pan-STARRS are assumed to have the same form as those for LSST, but have been adjusted to match the 
sensitivities given in Abdalla \etal (2008)
for a Pan-4 like scenario (Table \ref{tab:LSST_params}). 

From the ``true'' signals of the CMC, we thus created a catalog of realistic observations for LSST, for Pan-STARRS, and for Euclid combined with a proposed on-board $U+G$ optical channel.   

\subsection{Photometric Redshifts}
\label{method:photoz}

From the noisy observations, we then calculated a photometric redshift to each galaxy for various filter combinations. The photometric redshifts were estimated by comparing the simulated observed broadband photometry with a grid of the model SEDs from the CMC. While this creates a situation where our model SED templates perfectly match the ``reality'' of the CMC galaxies, and therefore
over-estimates the photo-$z$ quality, we feel this is an unavoidable approach. While some systematic effects of choice of SED templates are known (for example, the BC03 templates are thought to under-estimate stellar mass due to a poor treatment of the thermally pulsating asymptotic giant branch phase (Bruzual 2007)), most are not well understood. It would therefore be very difficult to realistically model the scatter and possible bias of the photometric redshift estimates as a consequence of our choice of model templates. We chose to instead focus on contributions to the photo-$z$ error resulting from random photon statistics and possible systematic instrument calibration errors, but acknowledge that the photo-$z$s given here may be slightly worse in reality because of imperfect SED templates.

To calculate the photometric redshifts, we used the SEDfit software package (Sawicki \& Yee 1998, Sawicki 2011 [in prep]). This software redshifted the CMC model spectra onto a grid of redshifts spanning $0 \leq z \leq 6$ in steps of 0.02 and attenuated them using the Madau (1995) prescription for continuum and line blanketing due to intergalactic hydrogen along the line of sight. It then integrated the resultant observer-frame model spectra through filter transmission curves to produce model template broadband fluxes. In order to match the model template fluxes to the simulated observations, the observed fluxes of each object were compared with each template in the grid by computing the statistic
\begin{equation}
\chi^2 = \sum_{i} {[f_{obs}(i) - sf_{tpt}(i)]^2\over\sigma^2(i)},
\end{equation}
where $f_{obs}(i)$ and  $\sigma$(i) are the observed flux and its uncertainty in the $i$th filter, and $f_{tpt}(i)$ is the flux of the template in that filter. The variable $s$ is the scaling between the observed and template fluxes, and can be computed analytically by minimizing the $\chi^2$ statistic with respect to $s$ giving
\begin{equation}
s = {\sum_{i} {f_{obs}(i)f_{tpt}(i) / \sigma^2(i)}\over{\sum_{i}{f_{tpt}^2(i) / \sigma^2(i)}}}
\end{equation}
(Sawicki 2002). For each object, the most likely redshift is determined by the smallest $\chi^2$ value over all the templates. Error bars are generated by refitting the object 200 times with slightly perturbed photometry and finding the range in which 68\% of the fits lie. 

\subsection{Figure of Merit}
\label{method:fom}

In order to objectively compare the different observational scenarios, we employ the DE Figure of Merit (FoM) proposed by the Dark Energy Task Force (DETF). This number is the inverse of the area of the 2-$\sigma$ uncertainty ellipse in the plane of the DE parameters $w_0$ and $w_a$. The FoM is thus a statement on the precision of the DE measurements, not necessarily the accuracy.

We used the iCosmo package (Refregier \etal 2008) to calculate the FoM for each of our survey scenarios assuming a flat cosmology with fiducial cosmological parameters of $(\Omega_m, w_0, w_a, h, \Omega_b, \sigma_8, n_s, \Omega_\Lambda)$ $=$ $[ 0.3, -0.95, 0, 0.7, 0.045, 0.8, 1, 0.7]$, an intrinsic ellipticity dispersion of 0.25, and 10 tomographic redshift bins. The calculations are done using only the weak lensing power spectrum which is summed over $10\leq\ell\leq20000$, and the $w_0-w_a$ uncertainty is marginalized over the other five parameters without any external priors.

To orient our comparisons, the Euclid Science Book (Refregier \etal 2010) states that the current FoM is on the order of 10, which is generated using WMAP observations combined with Baryon Acoustic Oscillation (BAO) and Type Ia supernova distance measurements, as well as a prior adopted in accordance with Big Bang Nucleosynthesis (Komatsu \etal 2009). This is the FoM value currently achieved by combining several available DE probes;  in contrast, in the rest of the paper we presenting FoM values attainable from weak lensing observations alone, without the inclusion of other probes available now or in the future. 

A FoM generated {\it{solely}} from a weak lensing survey using both space-based and ground-based observations is expected to be approximately 180 (see Table 4.1 in Euclid Science Book), over an order of magnitude greater than the current figure. However, values can vary depending on the parameters and methods used to estimate the expected FoM. For example, Amara \& Refregier (2007) obtain a FoM of only 50 for a survey with properties similar to what is expected with Euclid plus ground-based observations (20 000 deg$^2$ area, 35 gals/arcmin$^2$, median redshift of 0.9). Regardless, if all cosmological probes observable with Euclid are utilized then the FoM increases to $\sim400$, and well over 1000 with the use of external prior constraints derived from Planck.

It is informative to discuss how various parameters of the photometric redshift distribution affect the FoM. Amara \& Refregier (2007) have shown that the FoM is almost directly proportional to the number density of galaxies in the photometric redshift catalog. However there is a trade-off between a wide and a deep survey, as they show that the FoM also depends strongly on the median redshift of the photo-$z$ distribution (FoM $\propto z_m^{1.2}$). The figure of merit also degrades as the precision of the photometric redshifts decreases (FoM $\propto 10^{-1.69\sigma_z}$) and as the fraction of objects with catastrophic redshift errors ($F_{cata}$) increases (FoM $\propto 10^{-0.75F_{cata}}$). It is clear that accurate and precise photo-$z$s are important if a respectable FoM is to be obtained.

%%%%%%%%%%%%%%%%%%%%%%%%%%%%%%%%%%%%%%%%%%%%%%%%%%%%
%SECTION 3
%%%%%%%%%%%%%%%%%%%%%%%%%%%%%%%%%%%%%%%%%%%%%%%%%%%

\section[Results]{Results}
\label{results}

In this work, we used the above procedure to create a catalog of observations for each galaxy in the mock catalog, but limited ourselves to analyzing only those galaxies that have an $AB$ magnitude less than 24.5 in the $RIZ$ channel. We studied three options for a weak lensing survey: 1) using only the filters currently planned for the Euclid instrument ($RIZ$ shape channel, $Y, J$ and $H$), 2) the Euclid filters plus additional on-board optical filters $U$ and $G$, and 3) the Euclid IR filters ($Y$, $J$, $H$) plus ground-based observations from either Pan-STARRS ($grizy$) or LSST ($ugrizy$). The first scenario was done strictly for comparison and is not expected to yield usable photometric redshifts since it has no optical observations. The second scenario we divided into several sub-cases in which we examined the impact of different exposure times with the two optical filters and also the effect of not using the $RIZ$ shape channel for photometry. For the last case we preferentially used LSST for the ground based observations, since we found similar although slightly worse photometric redshift results using Pan-STARRS (see also Abdalla \etal 2008). A summary of the scenarios and results is presented in Table \ref{tab:scenarios}. The median redshift for all cases is 0.8.

\subsection{Euclid Alone}
\label{results:euclid}

\begin{figure}
\includegraphics[width=8cm]{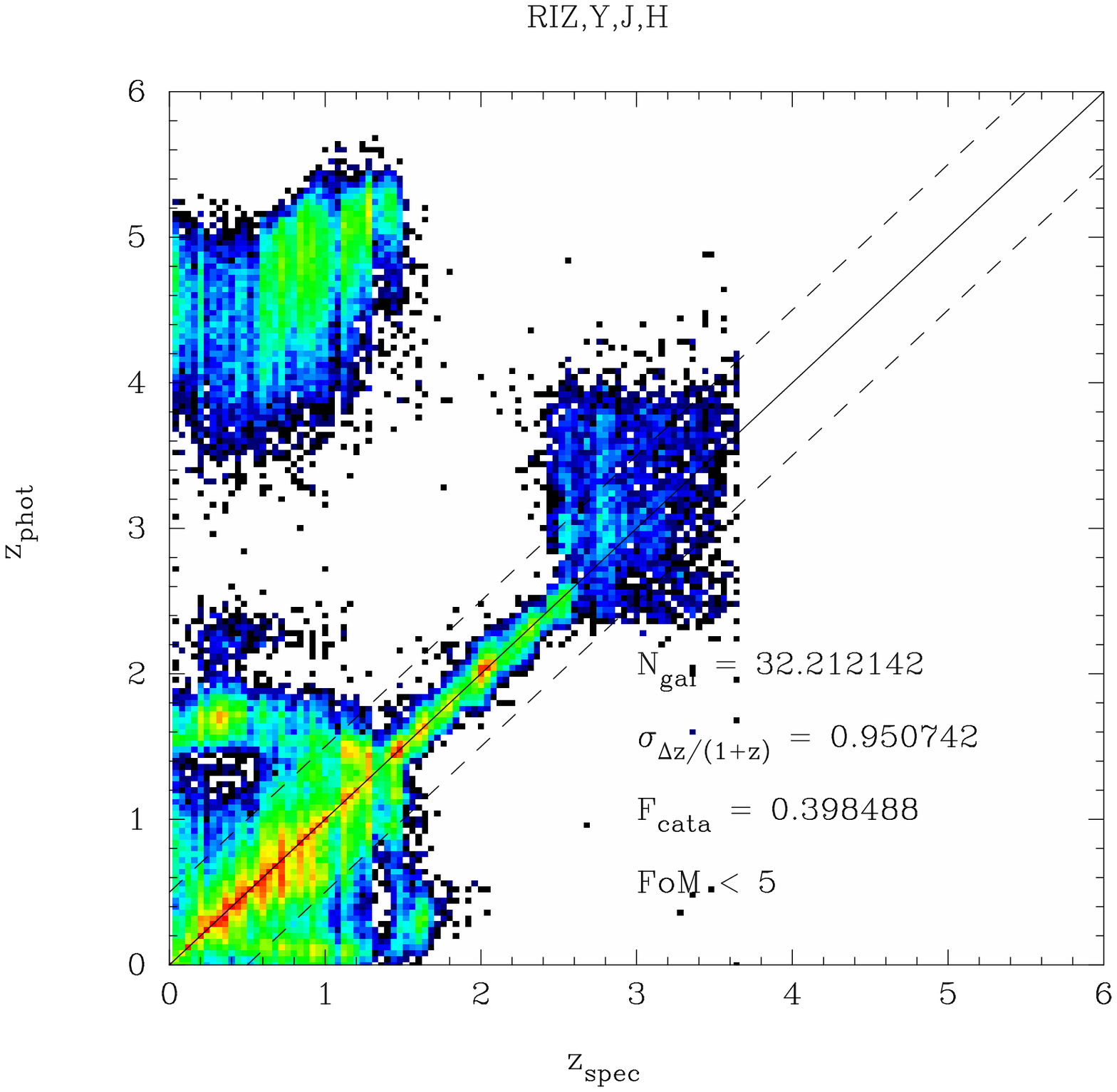}
\includegraphics[width=8cm]{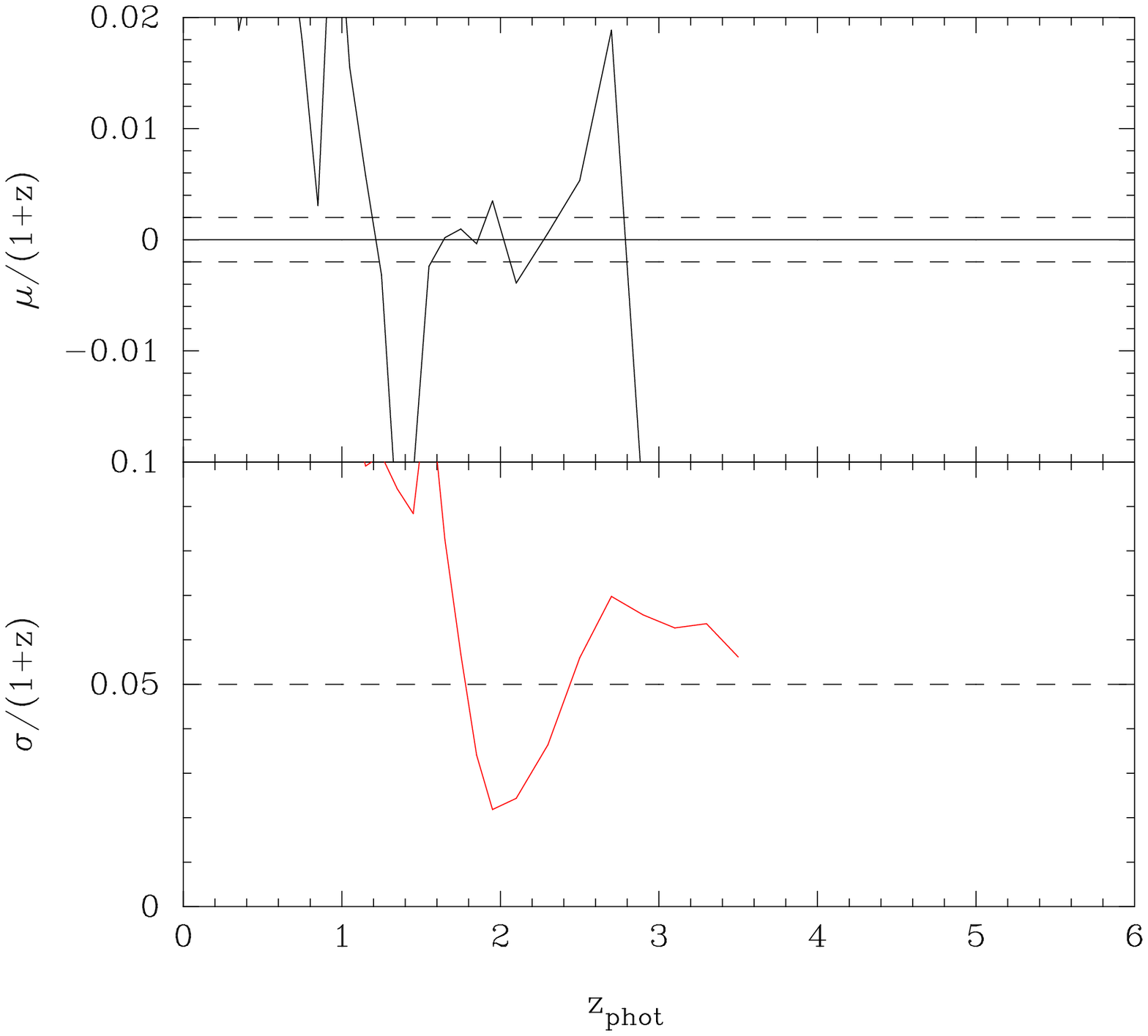}
\caption{\label{fig:Euclid_only}Left: Photometric redshift as a function of spectroscopic redshift using only Euclid's $RIZ,Y,J$ and $H$ band-passes. $N_{gal}$ is the number density of galaxies per arcmin$^2$, $\sigma_{\Delta{z}/(1+z)}$ is the overall standard deviation for all galaxies, and $F_{cata}$ is the fraction of catastrophic redshifts defined to be $\Delta{z} > 0.3$ (shown by the dashed diagonal lines). Right: Standard deviation ($\sigma$) and bias ($\mu$) of the photometric redshifts scaled by $1+z$ as a function of redshift; the bias does not include any corrections which may be possible through spectroscopic calibration. The dashed horizontal lines in the $\mu$ and $\sigma$ panels show the scientific requirements for a weak lensing survey. The FoM for this scenario is less than 5.}
\end{figure}

\begin{figure}
\includegraphics[width=8cm]{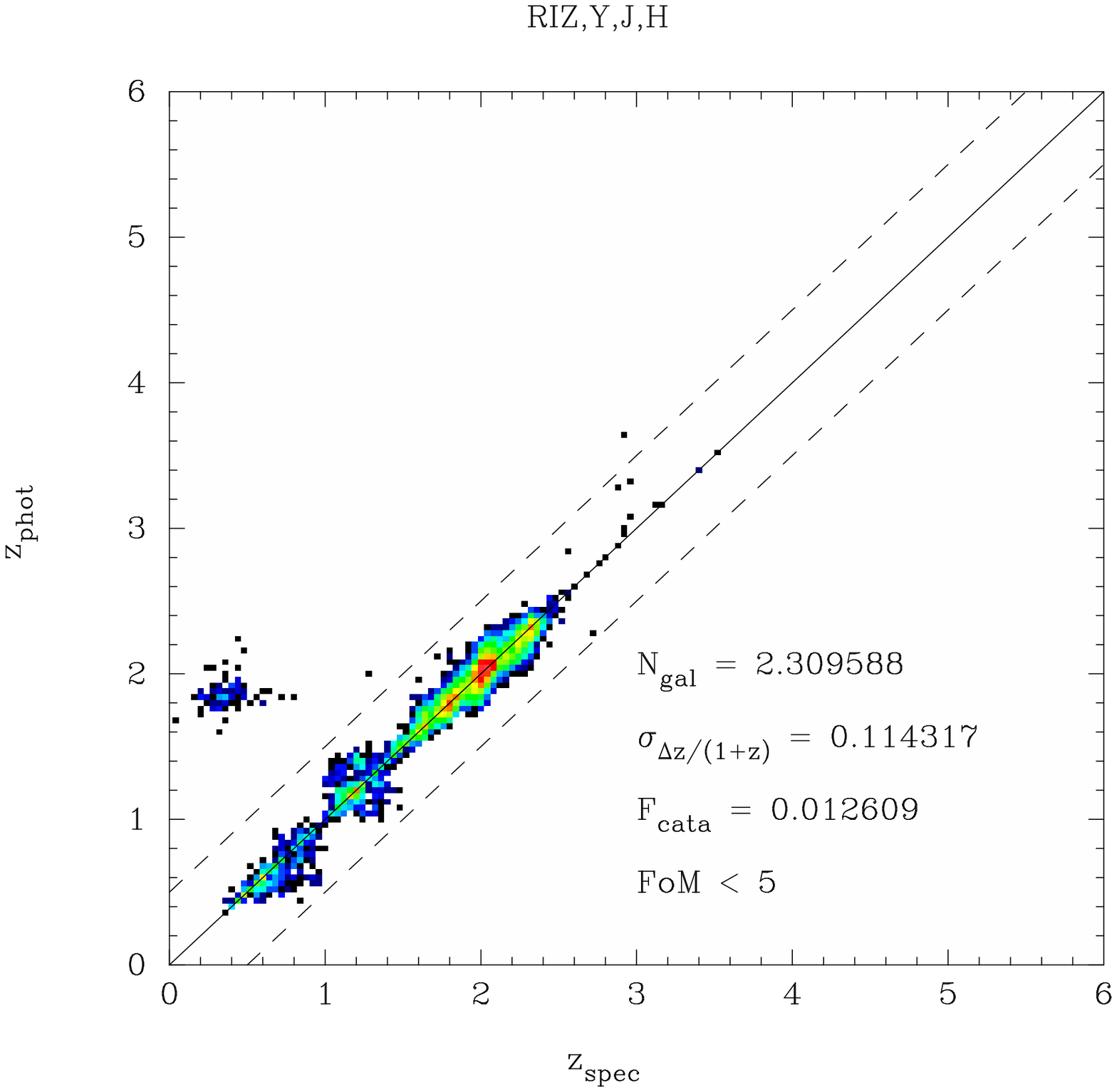}
\includegraphics[width=8cm]{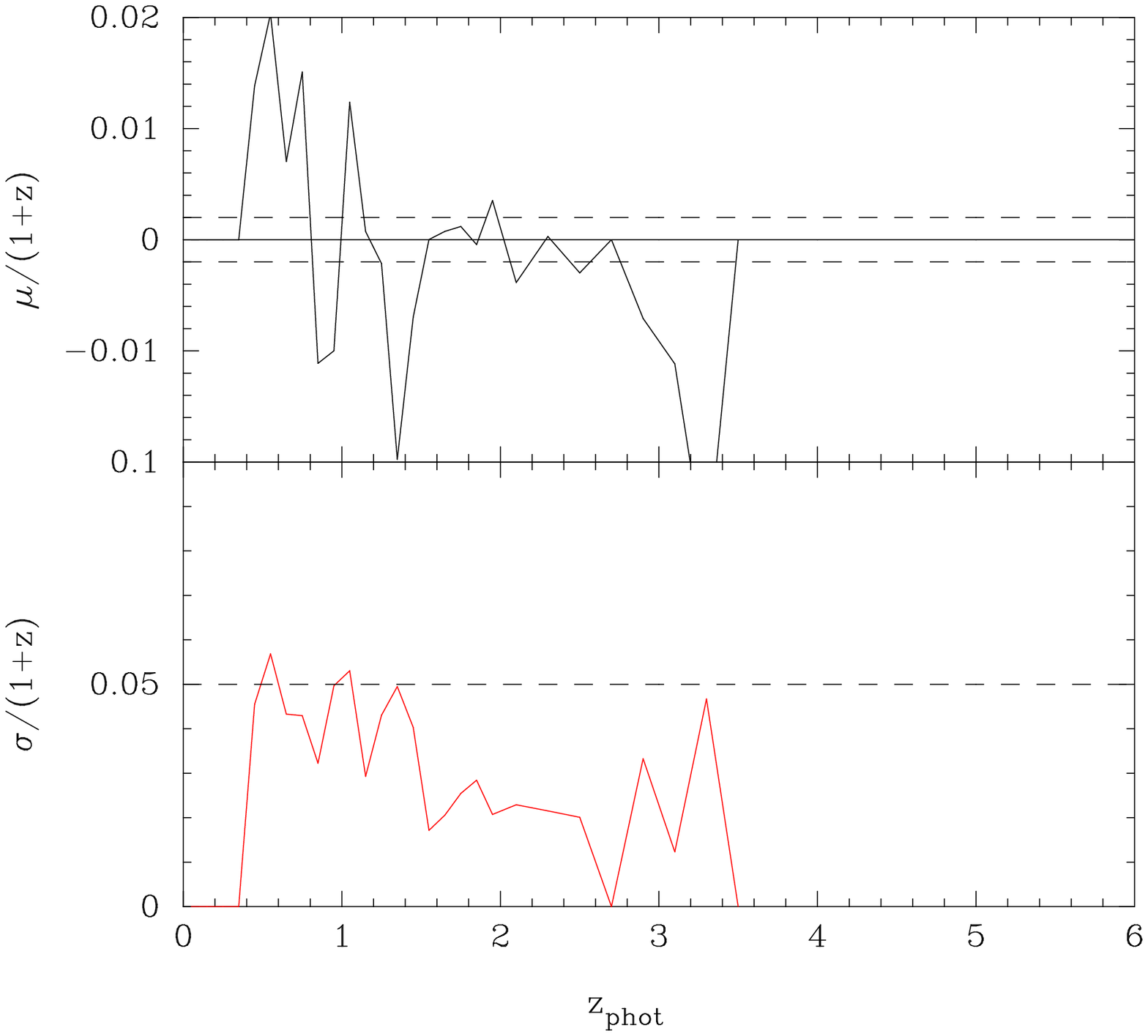}
\caption{\label{fig:Euclid_only_trust5}Same as Figure \ref{fig:Euclid_only} except culling galaxies which have a poorly constrained photometric redshift with error bars greater than 0.5.}
\end{figure}

The currently proposed strategy for Euclid has chosen to rely on other ground-based projects to obtain optical measurements for photometric redshift estimation. The optical wavelength observations are required in order to obtain accurate low redshift photo-$z$s by detecting the various breaks in a galaxy's SED as they appear in our observer frame. It is well understood that without any optical band-passes, virtually no constraints can be placed on the redshifts of low-$z$ galaxies. Thus, it is no surprise that the plot shown in Figure \ref{fig:Euclid_only} contains a large number of catastrophic redshifts as galaxies with $z\leq1$ are scattered upwards to higher redshifts. Note also that the standard deviation $\sigma_z\over{1+z}$ is well above the required level of 0.05 at nearly all redshifts and the FoM is less than the present-day value. Obviously, results can be improved by culling galaxies that have low-quality photometric redshift estimates, as shown in Figure \ref{fig:Euclid_only_trust5} where any galaxy with an uncertainty $\Delta{z_{phot}}$ greater than 0.5 has been removed. Removal of poorly constrained galaxies can be a trade-off, as it tightens up the spread of the photo-$z$s and thus raises the FoM, but it also reduces the number density of galaxies which acts to lower the FoM. However, in this scenario, culling of poorly fit galaxies results in the removal of almost all galaxies needed for weak lensing in the target range of $0.3\leq z \leq 2$, resulting in a poor FoM.

The point of this exercise is to emphasize that when it is said that Euclid will rely on ground-based observations, it is {\it{fully}} reliant in that a weak lensing DE survey will not be possible without optical photometry from other telescopes.

\subsection{Addition of an On-board $U$ and/or $G$ Channel}
\label{results:ug} 

\begin{figure}
\includegraphics[width=8cm]{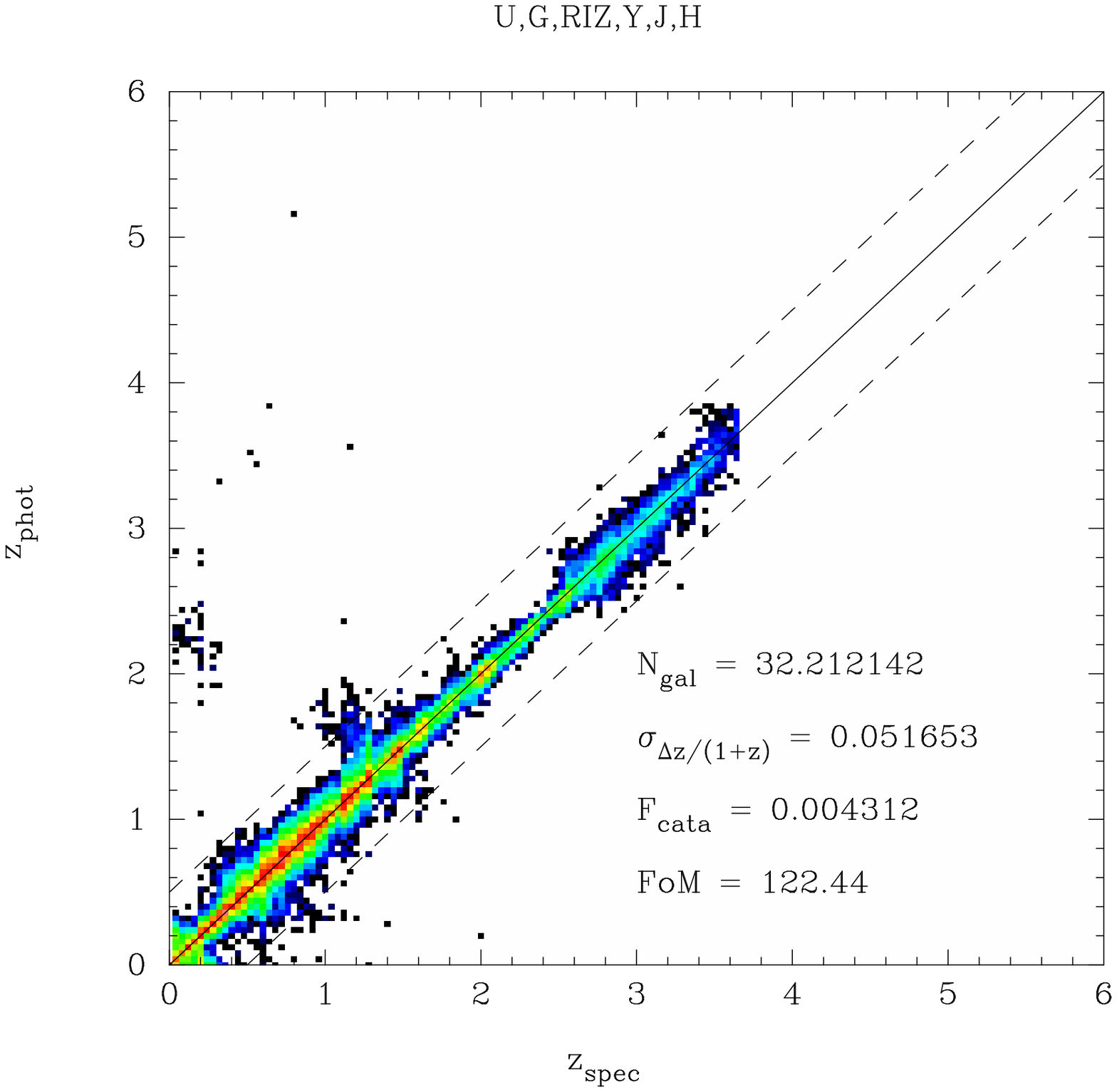}
\includegraphics[width=8cm]{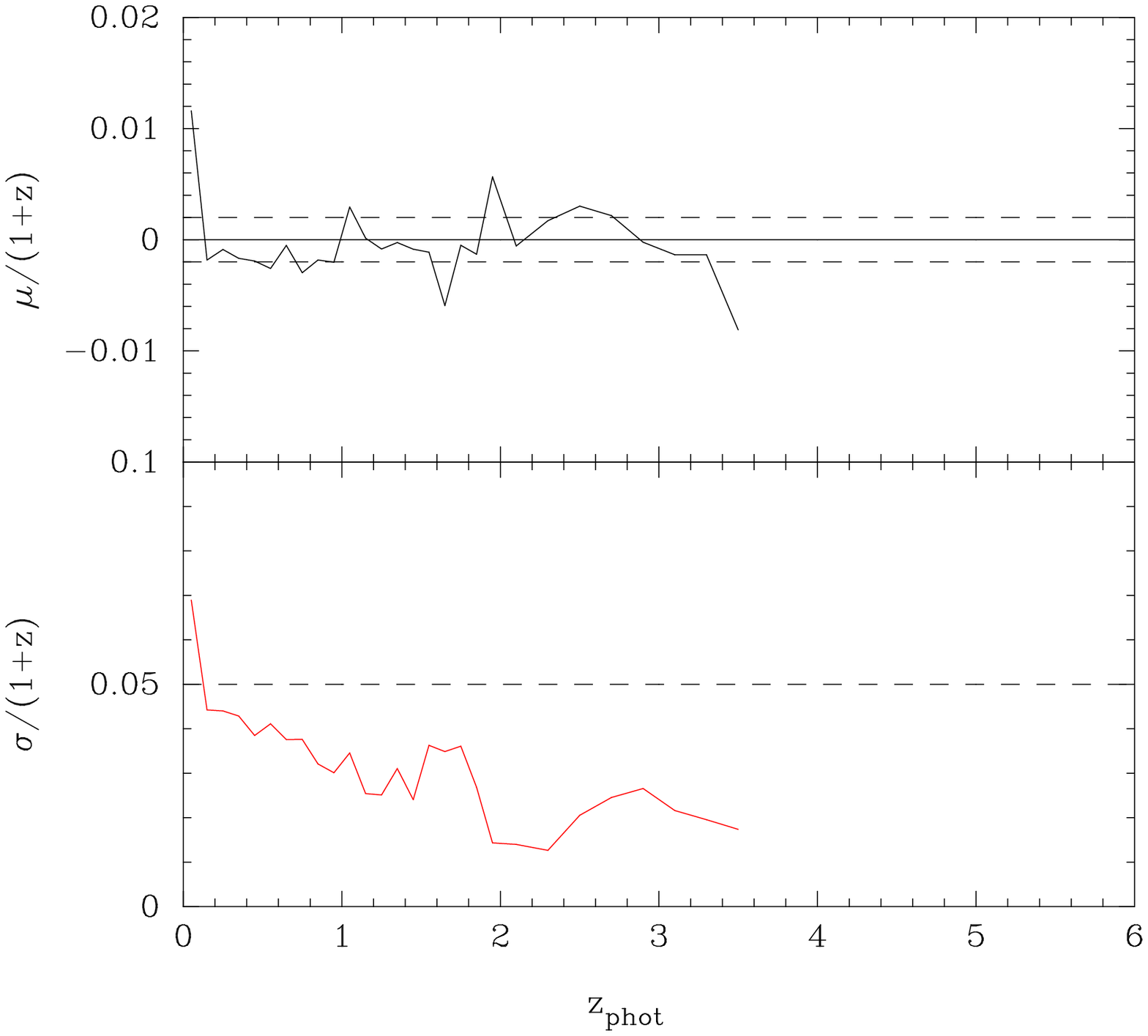}
\caption{\label{fig:Euclid_UG}Same as Figure \ref{fig:Euclid_only} except with the addition of $U$ and $G$ band-passes.}
\end{figure}

\begin{figure}
\includegraphics[width=8cm]{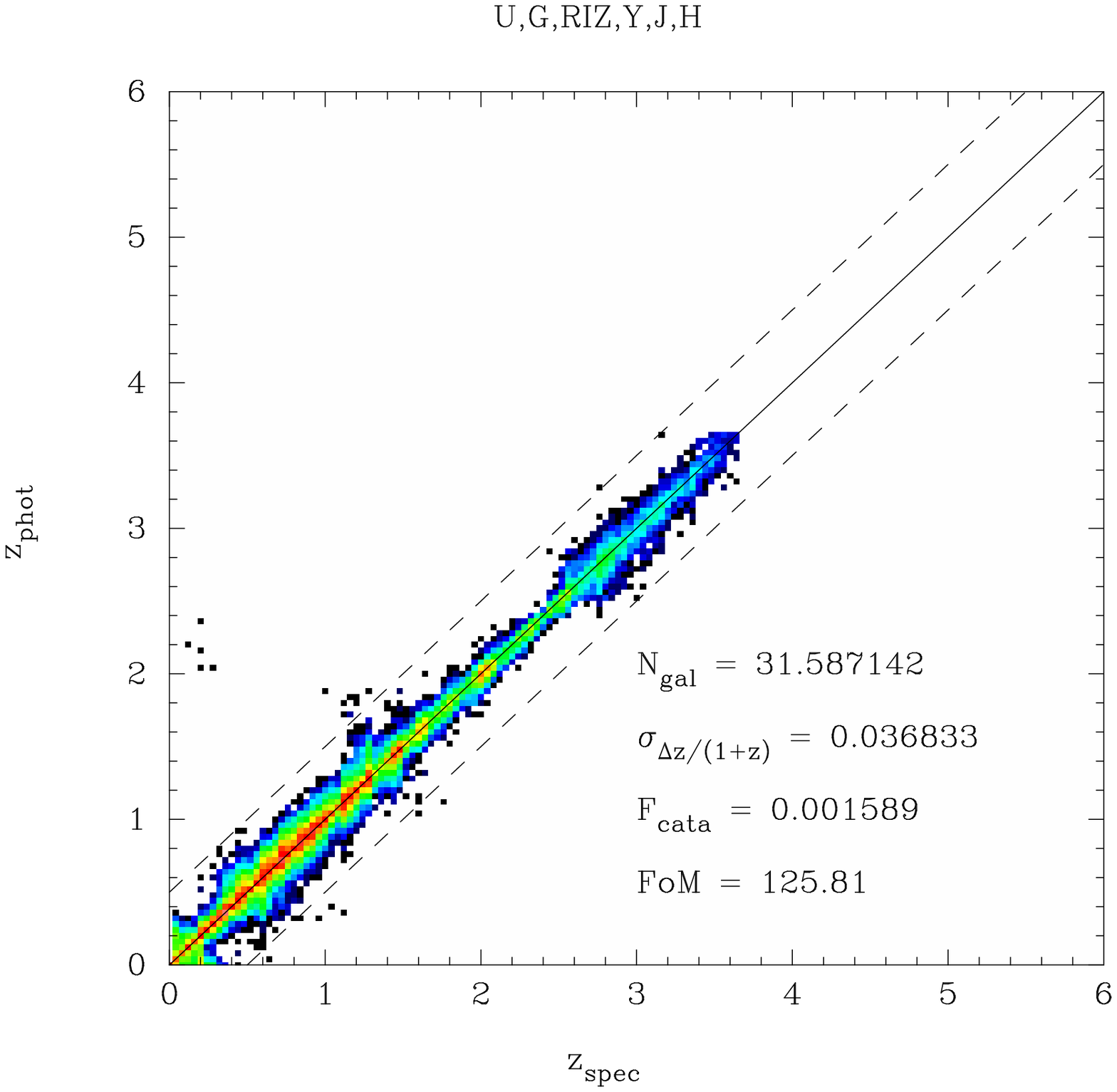}
\includegraphics[width=8cm]{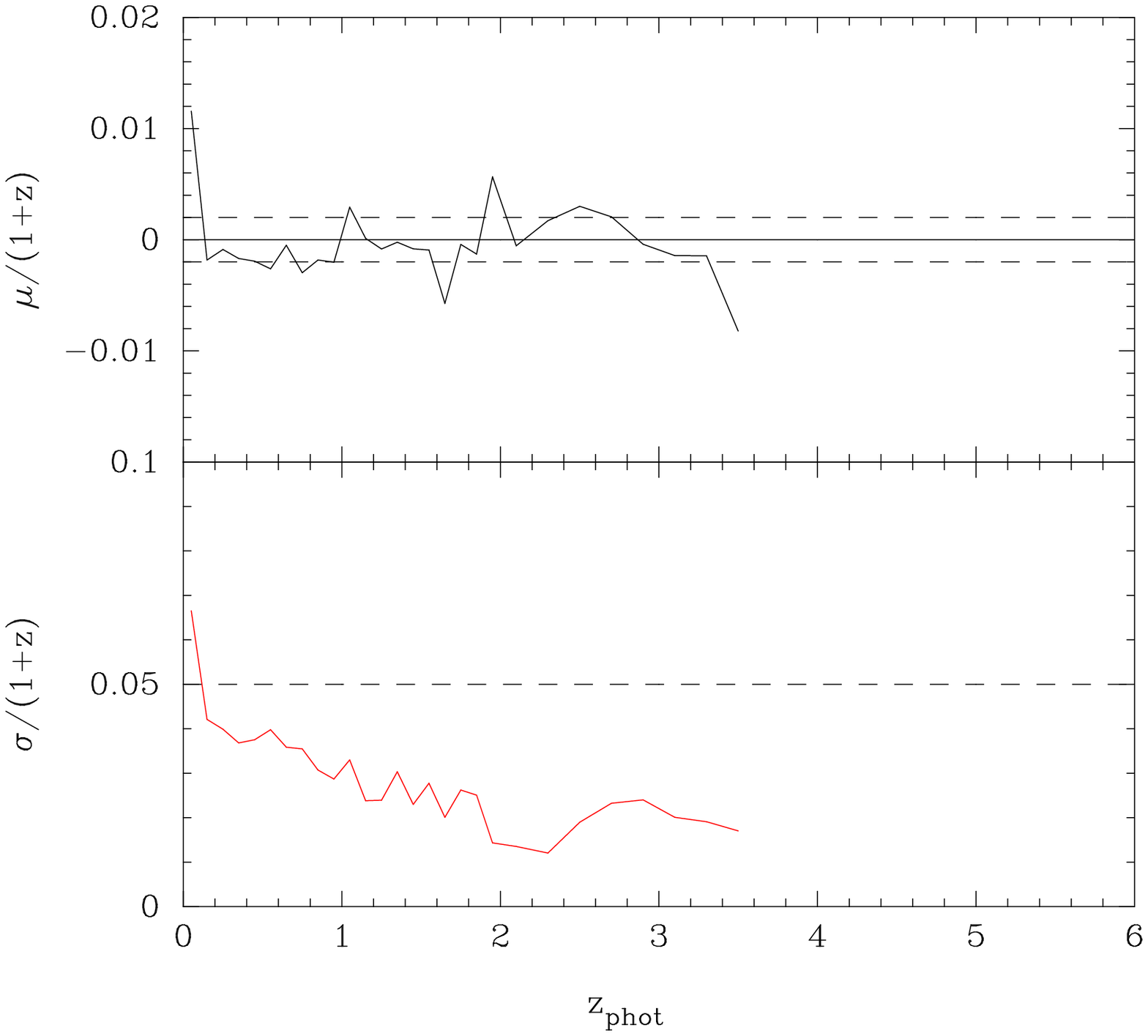}
\caption{\label{fig:Euclid_UG_trust5}Same as Figure \ref{fig:Euclid_only} except with the addition of $U$ and $G$ band-passes and culling galaxies with photo-$z$ error bars greater than 0.5.}
\end{figure}

Figure \ref{fig:Euclid_UG} shows the drastic improvement in photometric redshifts that can be found with the addition of two on-board optical band-passes to Euclid. The ability to discriminate low-$z$ galaxies from high-$z$ ones is invaluable. While the overall standard deviation is still rather high (${\sigma\over{1+z}}>0.05$) due to the number of catastrophic failures at low redshift, a simple culling of untrustworthy galaxies ($\Delta{z_{phot}}<0.5$) brings this down to below the required level as shown in Figure \ref{fig:Euclid_UG_trust5}. The right panels of Figures~\ref{fig:Euclid_UG} and \ref{fig:Euclid_UG_trust5} demonstrate that even if the overall standard deviation is high, the standard deviation as a function of redshift can still be below the required value in the key redshift range for weak lensing ($0.3 \leq z \leq 2$). The resulting FoMs of 122 and 126 for the raw and culled scenarios respectively make it clear that the addition of on-board optical band-passes could make Euclid self-sufficient for weak lensing. Although observations from other instruments could of course still be used, Euclid would no longer be completely reliant on them. 

\begin{figure}
\begin{center}
\includegraphics[width=8cm]{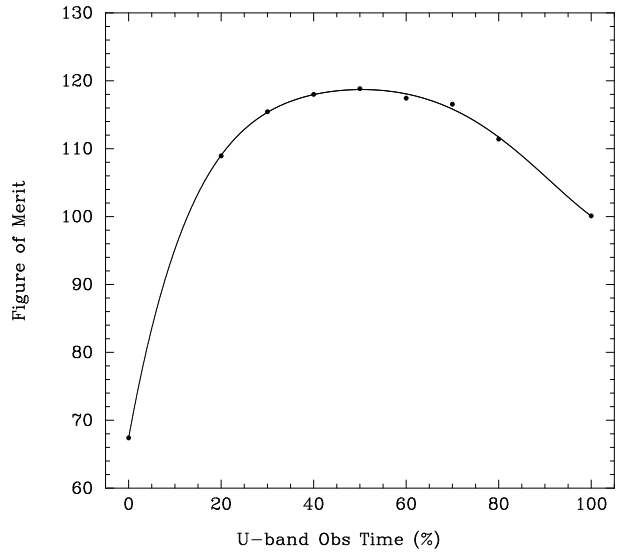}
\end{center}
\caption{\label{fig:ugcombo}Demonstrating the effects of time sharing between the $U$ and $G$ filters. The plot shows how the FoM changes as the percentage of time spent observing in $U$-band increases. The $G$ band observing percentage is 100 minus the $U$ band percentage. The total observing time is 542s. The solid line is a fifth order polynomial fit to the data points.}
\end{figure}

\begin{figure}
\includegraphics[width=8cm]{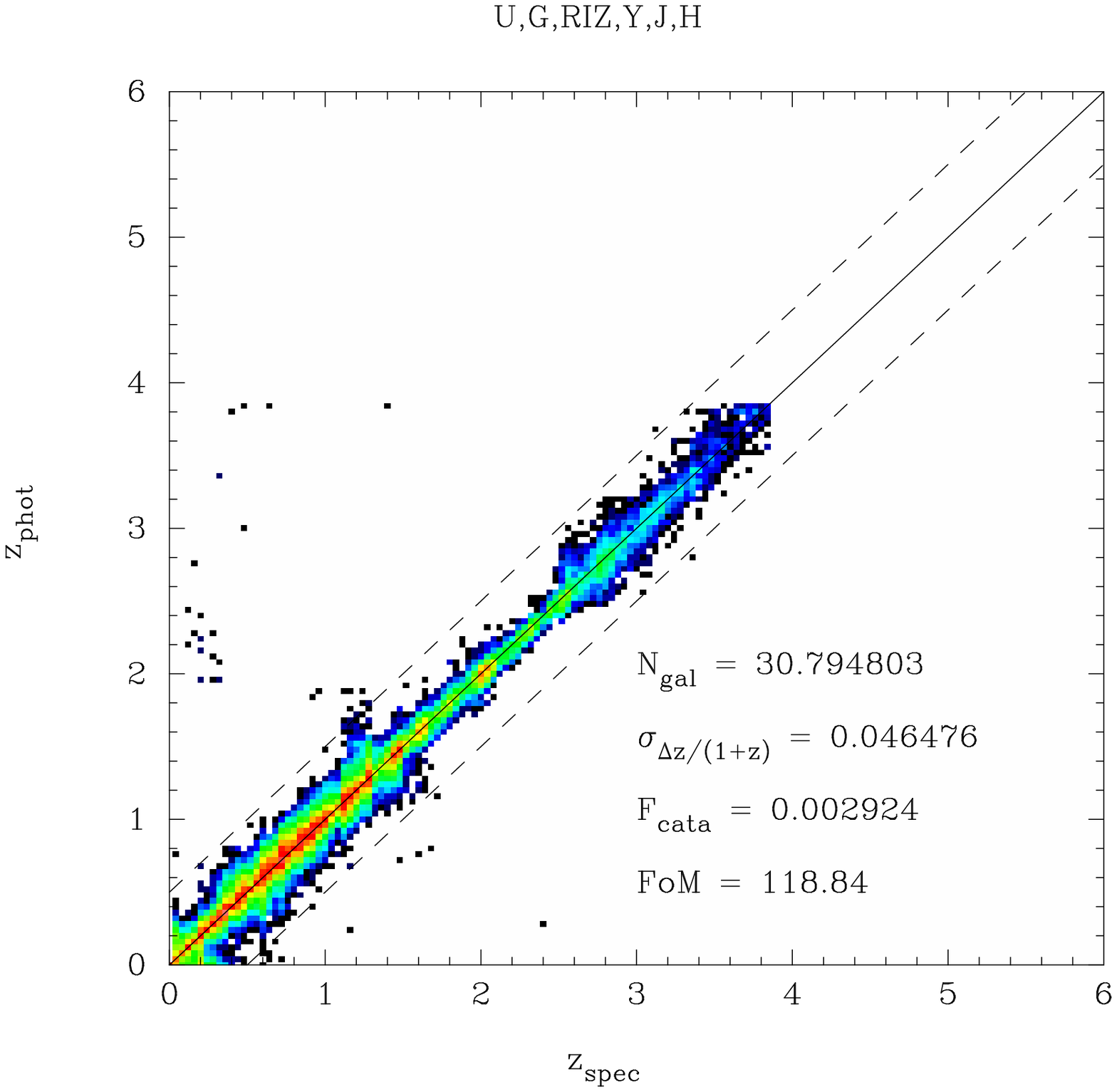}
\includegraphics[width=8cm]{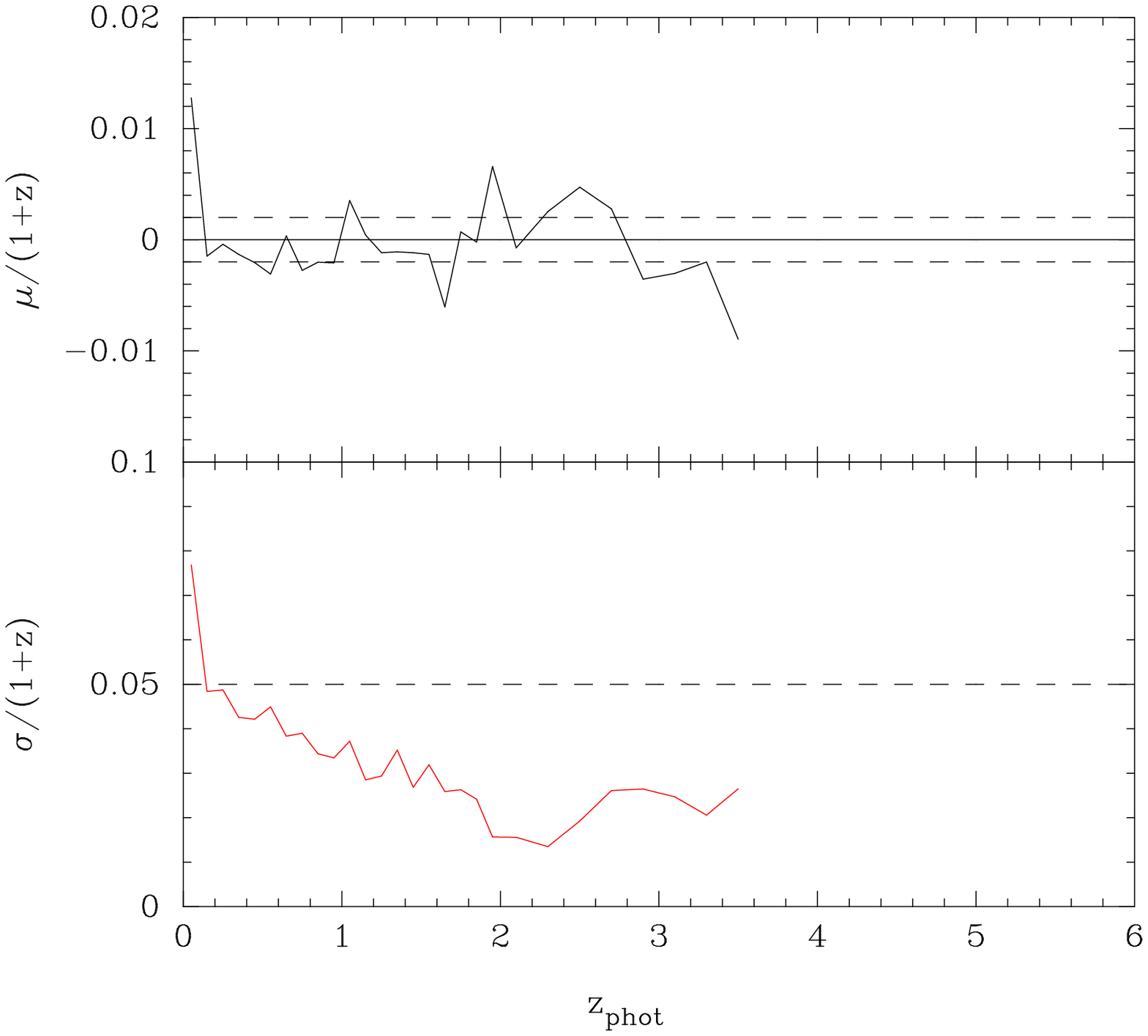}
\caption{\label{fig:combozz}Same as Figure \ref{fig:Euclid_only} except using band-passes $U,G,RIZ,Y,J,H$ where the $U$ and $G$ bands each observe for only 271s (50\% of the $RIZ$ band) and galaxies with photo-$z$ error bars greater than 0.5 have been culled.}
\end{figure}

In the above scenario, both the $U$ and $G$ filter feed dedicated detectors and so have the maximum exposure time available ($\sim$ 500s, the same as the $RIZ$ shape channel).
If the two optical bands cannot observe simultaneously, but instead have to share observation time as might be the case if a single detector plus a filter exchange mechanism were used, this will have a negative impact on the photometric redshift estimations. Figure \ref{fig:ugcombo} shows that the optimal time-sharing arrangement would be approximately an even division of time. The FoM has a maximum at near 50\% observing time in each band of 118.84. The photometric redshift distribution of this best-case scenario is shown in Figure \ref{fig:combozz}.

The FoM stays at roughly the same level ($115 < FoM < 119$) until the observing time percentage drops below 30\% in either band. At the extreme ends of sharing scenarios, it is clear that $U$ band observations are more critical than $G$ band observations, as the FoM with nothing but $G$ is 67.4, much less than the FoM = 100.1 for solely $U$ band observations. The larger FoM comes from the better constraints the $U$ band can place on the lowest redshift galaxies. In order to determine if two optical filters are absolutely necessary, we also tested a scenario with a broad filter that combined the $U$ and $G$ wavelengths which resulted in a FoM of 87.86.
While being able to estimate the photo-$z$s of low-redshift galaxies better than just the $G$ band, the broadness of this merged filter led to ambiguity as to where the 4000\AA\ break falls within the filter, and hence it was not able to break the degeneracy between low-$z$ and high-$z$ objects as well as simply the $U$ filter.
These results show that both a $U$ and a $G$ filter are required for optimal photometric redshifts. Note, however, that using just the $U$ filter gives a fairly adequate FoM of $\sim$100 which, while not optimal, may prove to be the best compromise between the instrument's weight and complexity and the best obtainable DE constraints.

As a final scenario for Euclid, we tested the effects of splitting the broad shape channel $RIZ$ filter into two separate filters, hereafter called $R$ and $Z$ and adding $U$ and $G$ filters as well. In this scenario, all the filters feed one detector and would be mounted on a filter wheel. Euclid's planned total observing time per dither is fixed at $\sim$700s, which was divided among the $U, G, R$ and $Z$ filters allowing ten seconds to account for the time taken to change filters. We found that using four filters ($U, G, R,$ and $Z$) in this finite amount of time was not beneficial as the short observing time increased the signal to noise ratio. The best case we found used three filters ($U, R,$ and $Z$) with the observing time split roughly evenly between the three although slightly favoring the $U$ band (40\%, 30\% and 30\% of the observing time in the $U, R,$ and $Z$ bands respectively). This layout yields a FoM of 105, lower than the scenarios with one broad shape channel and two optical bands, but slightly higher than one broad shape channel and the $U$ band alone. However this figure is an upper limit at best, as it is uncertain if accurate shape measurements will be attainable in the $R$ or $Z$ band with this little observing time. If the shape channel were to be split up, a different survey strategy that allows more observation time per object may be preferable in order to increase the $S/N$ in the observations, both for photometry and shape measurements  

\subsection{Euclid's IR plus Ground-Based Observations}
\label{results:LSST}

For comparison, we now show what is expected to be obtainable with the use of ground-based telescopes plus the IR bands from Euclid. In these scenarios, all surveys are assumed to overlap completely and cover the same 20 000 deg$^2$ area. This yields a best-case result and if the overlap between Euclid's space-based survey and ground-based surveys turns out to be smaller, then the FoM would be negatively affected. 
In fact, the FoM scales linearly with the survey area (Amara \& Refregier 2007) so that if only half of the space-based survey overlaps with the ground-based component then the expected FoM will also be cut in half.

\begin{figure}
\includegraphics[width=8cm]{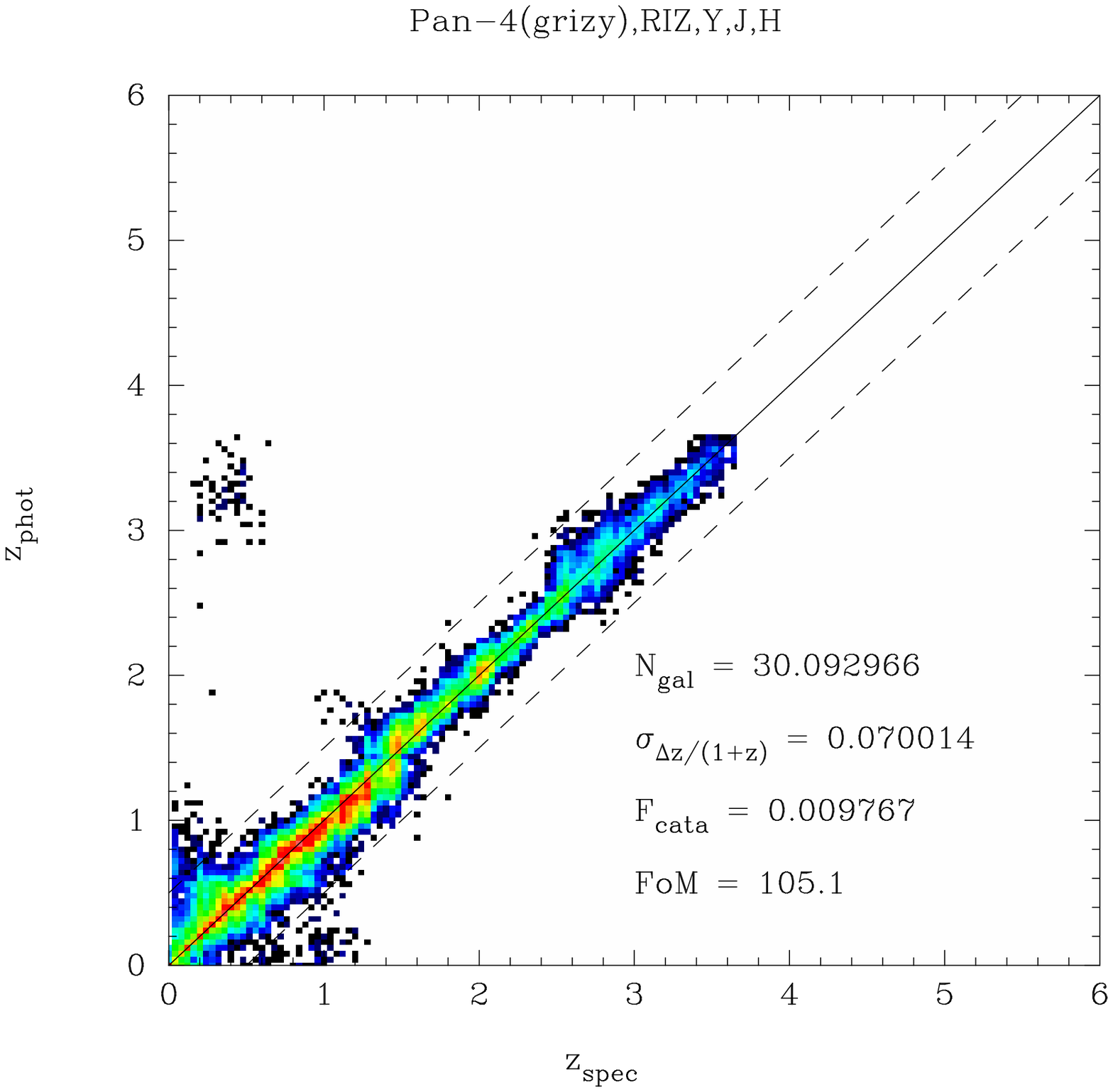}
\includegraphics[width=8cm]{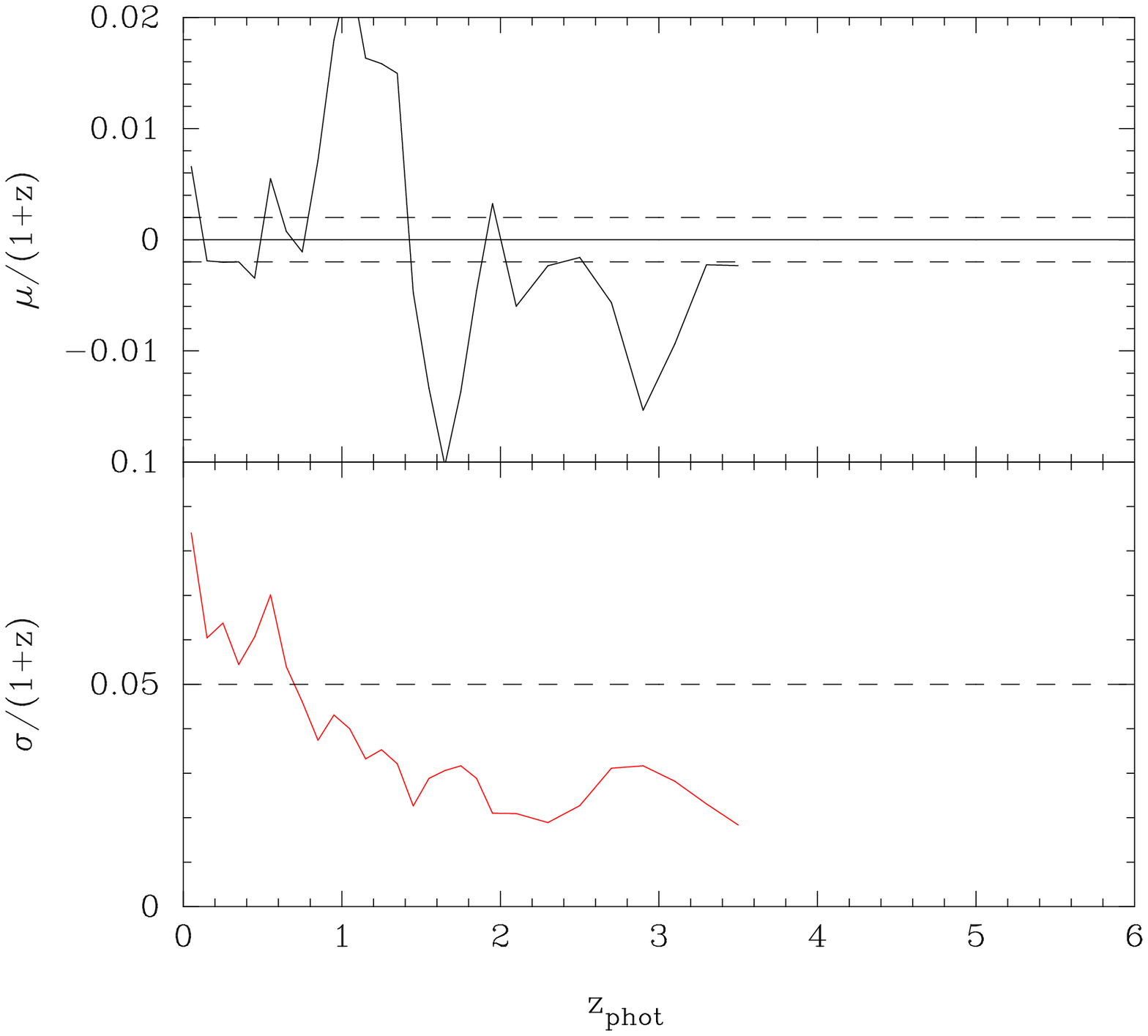}
\caption{\label{fig:Pan4}Same as Figure \ref{fig:Euclid_only} except using band-passes from both Euclid and Pan-STARRS with all four mirrors and culling galaxies with photo-$z$ error bars greater than 0.5.}
\end{figure}

\begin{figure}
\includegraphics[width=8cm]{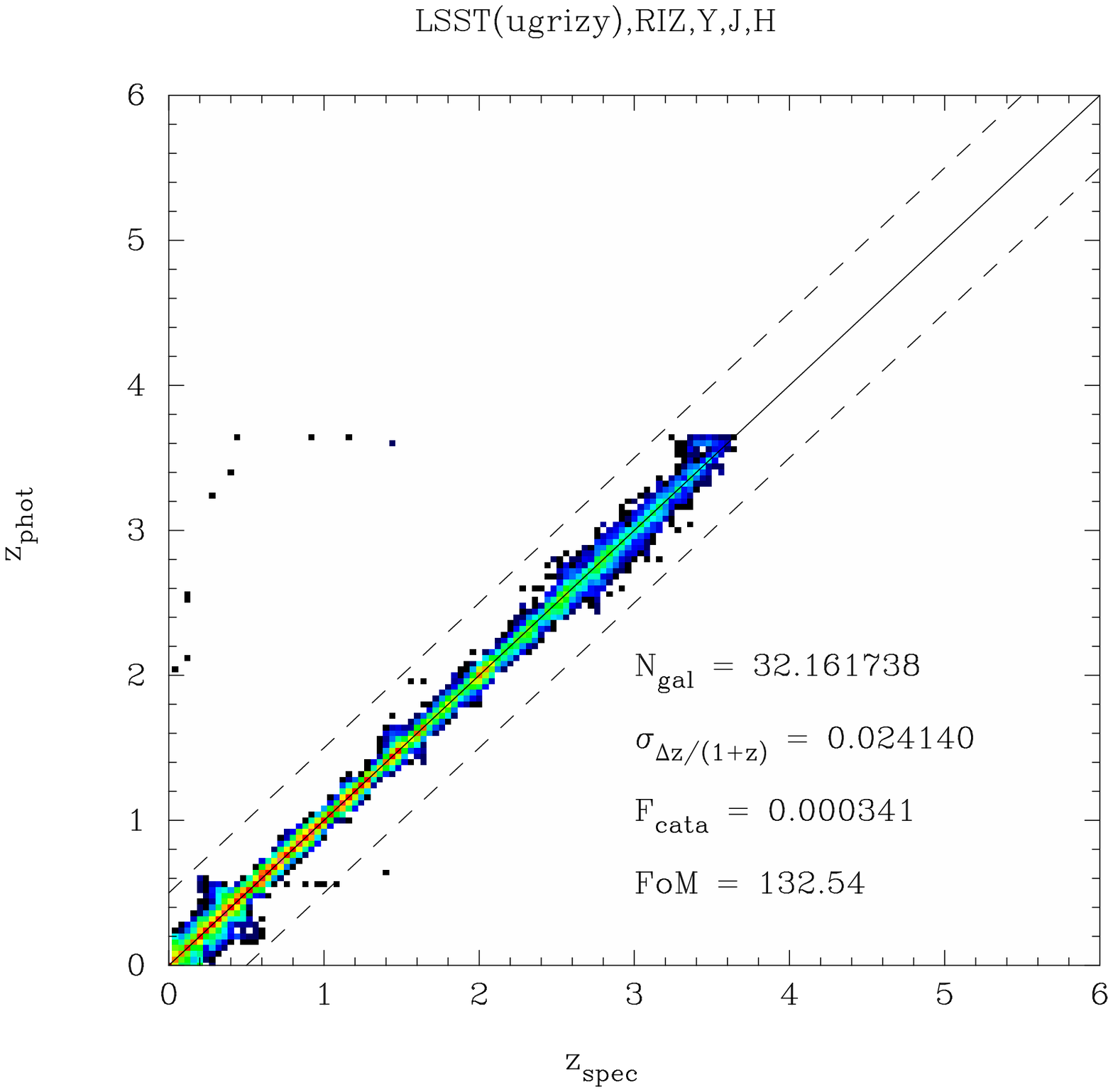}
\includegraphics[width=8cm]{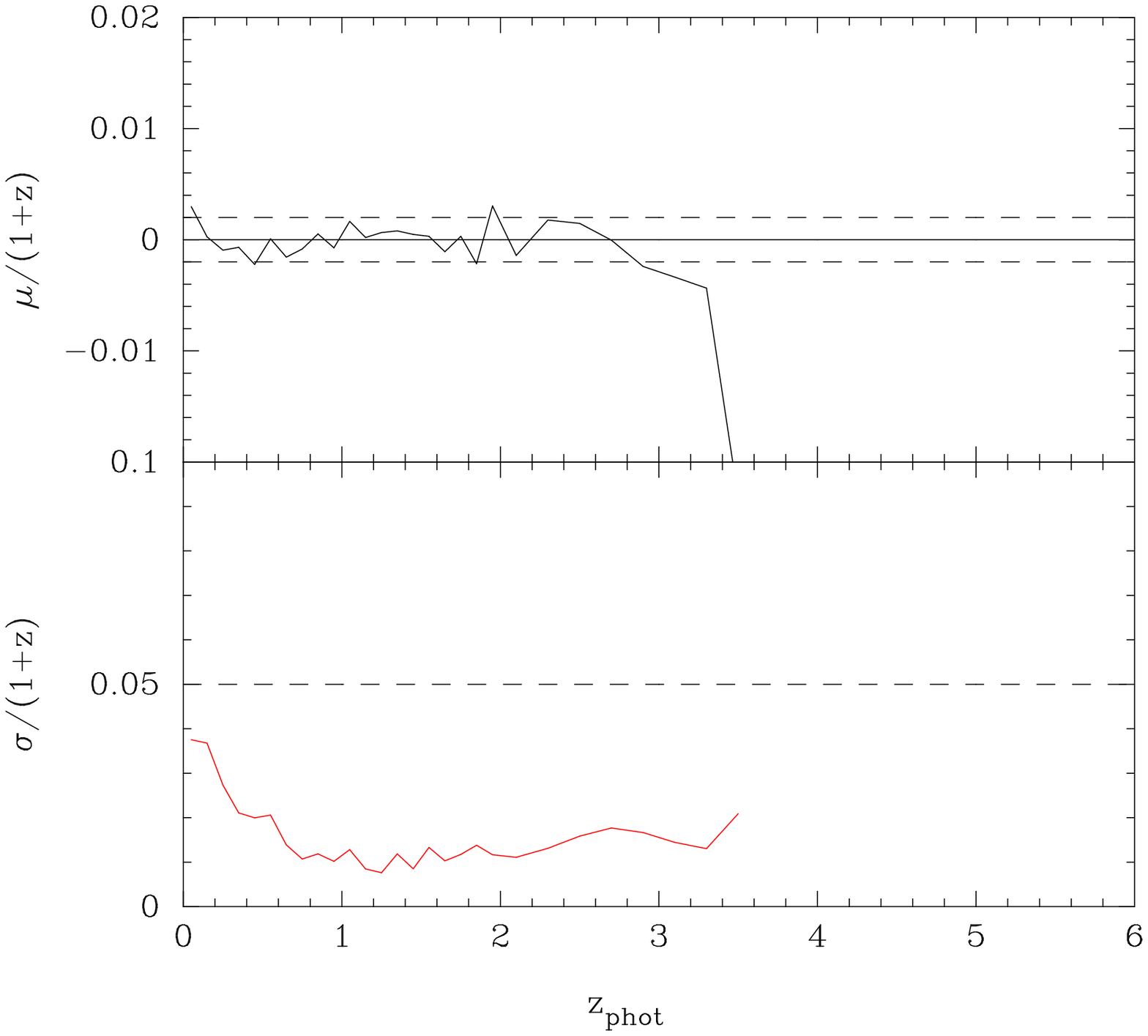}
\caption{\label{fig:LSST}Same as Figure \ref{fig:Euclid_only} except using band-passes from both Euclid and LSST and culling galaxies with photo-$z$ error bars greater than 0.5.}
\end{figure}

Figures \ref{fig:Pan4} and \ref{fig:LSST} show the results for Pan-4 and LSST respectively. The Pan-4 scenario is slightly worse than the Euclid+$UG$ case, while the LSST results are slightly better due to its deeper observations relative to Pan-STARRS. We will hence use the LSST observations for all future discussion. Abdalla \etal (2008) note that in order to obtain reliable photo-$z$s, shallower surveys such as DES or Pan-STARRS are not well matched to the Euclid survey, and show similar effects on the FoM due to the increased photo-$z$ scatter from these shallower surveys.

In the interest of establishing a time frame for the desired lensing results, we investigated how long it would take LSST plus Euclid to reach the same FoM as the Euclid plus on-board optical scenario, and found that LSST needs to observe each galaxy at least 45 times to have the same FoM (126) as Euclid plus optical filters (approximately 6 months of observations with LSST). After this point, the FoM will increase as LSST makes more and more observations. We now have a simple means of estimating how many LSST observations are needed if Equation \ref{eqn:LSSTerr} turns out to be overly generous in reality. For example, if the LSST photometric errors turn out to be twice as large as predicted, then four times as many observations (\ie 180 observations or $\sim2$ years) are needed to match the Euclid plus on-board optical FoM. 

We also studied two scenarios that could detrimentally affect the LSST observations: random photometric  zero-point errors, and systematic zero-point offsets between ground and space observations.

\begin{figure}
\includegraphics[width=8cm]{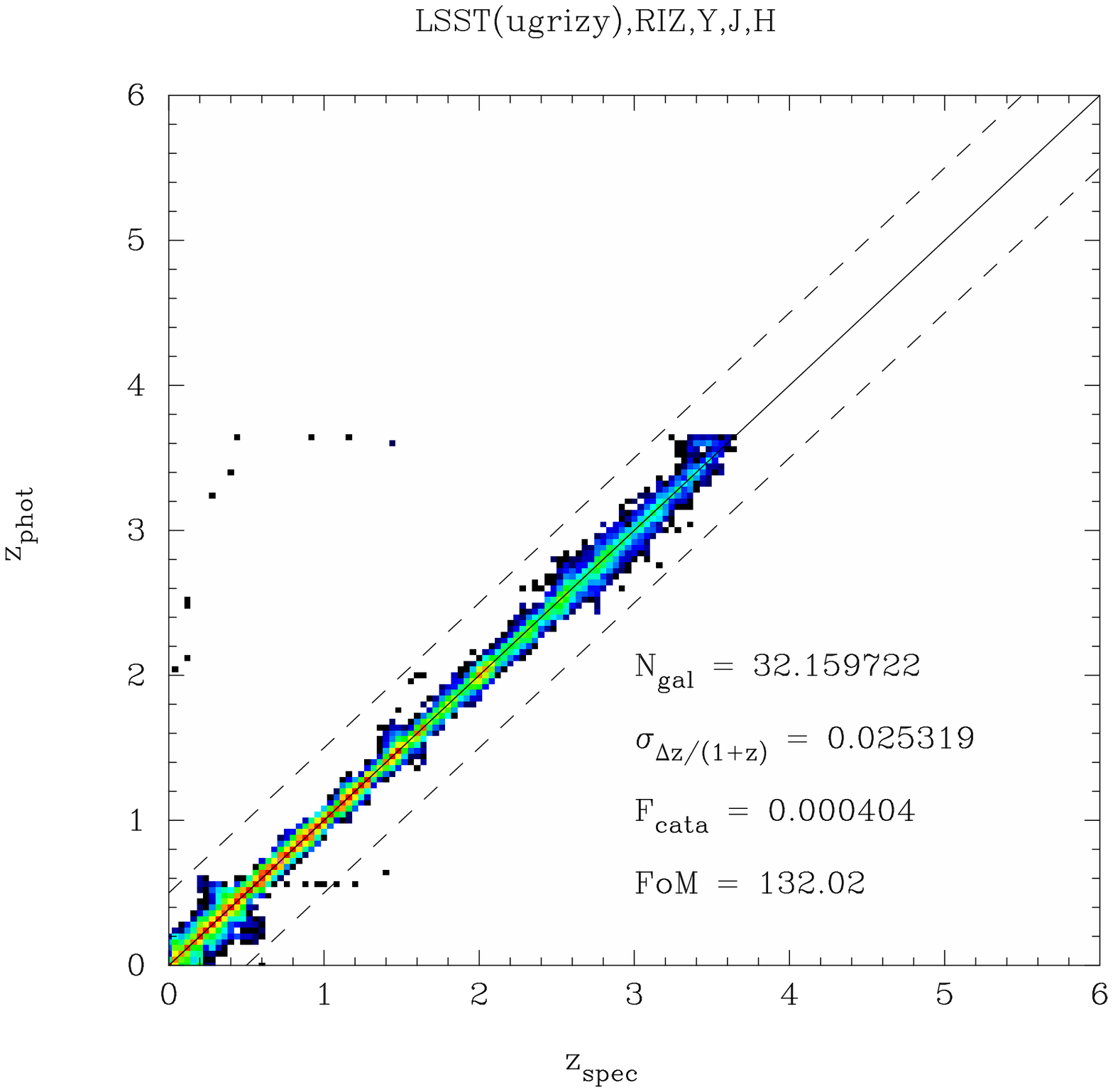}
\includegraphics[width=8cm]{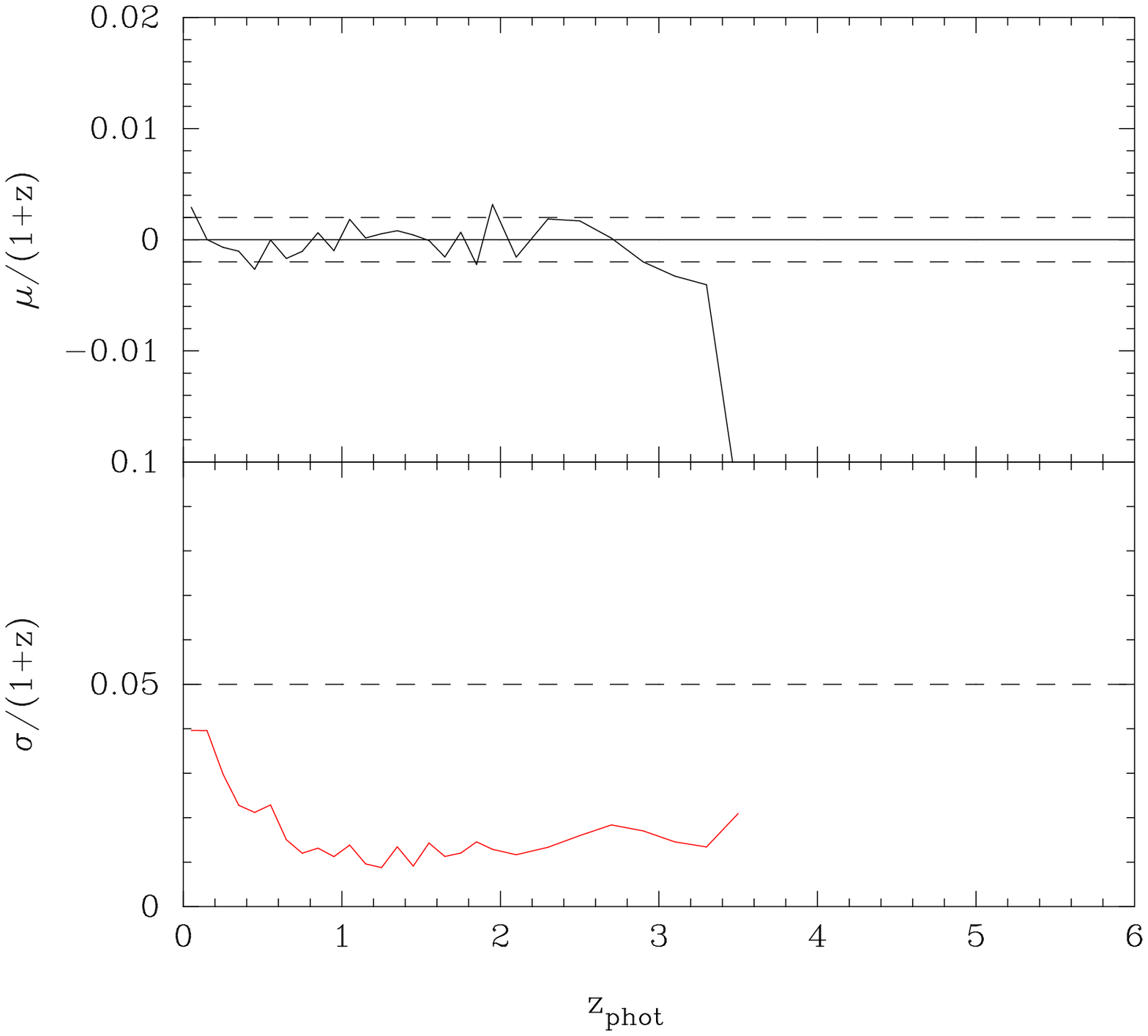}
\caption{\label{fig:LSST_offset}Same as Figure \ref{fig:LSST} except with random Gaussian errors added to simulate zero-point magnitude errors.}
\end{figure}

The error formulation given in Equation \ref{eqn:LSSTerr} for LSST specifically does not include any terms for mis-calibration of the zero-point (ZP) magnitudes from field to field. To study how any field-to-field ZP errors could negatively affect the photo-$z$s, we use the Canada-France-Hawaii-Telescope Legacy Survey (CFHTLS; Erben \etal 2009, Hildebrandt \etal 2009) as a reference point. Each pointing of the LSST could have slight errors associated with the ZP calibration and so we divided our observational catalog into $\sim2000$ fields, which is approximately the number of fields required for LSST to cover Euclid's 20 000 deg$^2$ survey. In each field we add a random Gaussian offset to the observed magnitudes in the various filters, where the standard deviations are similar to those found for CFHTLS (Hildebrandt, private communication) and are given in Table \ref{tab:LSST_offset}. This has the effect of increasing the spread in the photometric redshifts (shown in Figure \ref{fig:LSST_offset}) and thus slightly decreasing the FoM (though not significantly) to 132. The small deviations expected from field-to-field ZP errors are for the most part dominated by the random photometric errors. Additionally, the ZP errors are independent and thus add in quadrature with the photometric errors, leading to only slight effects in the best-fitting photometric redshifts. Their contribution is therefore almost negligible and we conclude that field-to-field ZP offsets in ground data are not likely to be an issue in DE weak lensing surveys.  

\begin{figure}
\begin{center}
\includegraphics[width=8cm]{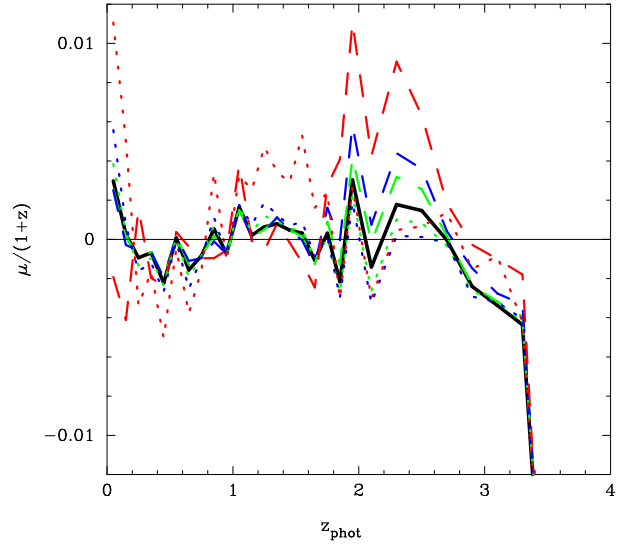}
\end{center}
\caption{\label{fig:mu}Effects of a systematic offset between the zero-point magnitudes of Euclid and LSST. The solid black curve is the ideal case of no offset, and the green, blue, and red curves show the result of increasing offsets of 0.01, 0.02 and 0.05 magnitude respectively. The dashed lines indicate that the four Euclid bands are offset by a negative amount relative to LSST magnitudes, while the dotted lines indicate a positive offset.}
\end{figure}

There is also a possibility that the LSST ZP magnitudes could be systematically offset from the ZPs derived for the Euclid instrument. We found that this has the effect of worsening the photometric bias $\mu_z$, specifically at redshifts greater than $\sim1.5$ when the 4000\AA\ break starts to fall between the LSST filters and Euclid's IR filters. Figure \ref{fig:mu} demonstrates how this effect worsens as the systematic offset increases. The steep drop-off in bias (present in nearly all the figures) above redshift 3 is a result of our brightness restriction, leading to small numbers of galaxies at this redshift and a bias towards galaxies that are erroneously bright in the $RIZ$ bandpass due to photometric errors. The dropoff should not be confused with a failure of LSST, as the sharp decline in bias is also seen in Figure \ref{fig:Euclid_UG}.

The rise in photo-$z$ bias can have detrimental effects to the weak lensing analysis, which requires the central redshift of each tomographic redshift bin to be known to better than 0.002(1+$z$). While many photo-$z$ codes can correct for systematic offsets between band-passes (\eg Ilbert \etal 2006, Coe \etal 2006), this requires spectroscopic redshifts for comparison. This highlights the importance of a spectroscopic redshift survey in order to properly calibrate the photometric redshifts. Such a survey is not without its own difficulties in that it has to overcome cosmic variance (van Waerbeke \etal 2006) and selection biases in order to obtain a fully representative sample. The calibration of any systematic ZP offsets might be greatly aided if the instruments shared similar band-passes, such as $U$ or $G$, and avoided entirely if space-based surveys could be independent.

%%%%%%%%%%%%%%%%%%%%%%%%%%%%%%%%%%%%%%%%%%%%
% SECTION 4
%%%%%%%%%%%%%%%%%%%%%%%%%%%%%%%%%%%%%%%%%%%
\section[Conclusions]{Conclusions}
\label{conclusions}   

In this work, we have used the currently proposed Euclid design to study how future space-based weak lensing missions might be able to estimate photometric redshifts independently, \ie without the use of complimentary ground-based observations.
We found that the addition of two or even one optical band-passes to Euclid could greatly improve the fidelity of photometric redshifts the telescope can attain by itself. If the $U$ and $G$ filters are added, the constraints that Euclid will be able to place on dark energy from weak lensing (FoM = 119--126) are comparable to those using a combination of Euclid and ground-based LSST observations (FoM = 132). Additionally, quite acceptable dark energy constraints can be obtained if only the $U$ bandpass is added to the baseline Euclid design (FoM = 100).

In their present form, to fullfil their weak lensing goals missions such as Euclid and WFIRST must rely on external ground-based observations. Including such ground-based observations entails many of the difficulties of combining two very large and different data sets, including, but not limited to, logistical complications, mis-match or potential delays in construction timescales, changes is planned survey designs, or data access limitations. 
Furthermore, if the survey area of the space-based observations does not overlap entirely with that of the ground-based survey, then the FoM will be negatively affected. For example, if Euclid and LSST only share 10,000deg$^2$, then the FoM obtainable by combining the observations will be only $\sim$66 instead of $\sim$132 in the case of full overlap.
The addition of on-board optical imaging through two or even one filter would avoid most of such complications. It would allow future space-based instruments to meet their scientific requirements for weak lensing without having to risk relying on external data.


%% Included in this acknowledgments section are examples of the
%% AASTeX hypertext markup commands. Use \url without the optional [HREF]
%% argument when you want to print the url directly in the text. Otherwise,
%% use either \url or \anchor, with the HREF as the first argument and the
%% text to be printed in the second.

\acknowledgments

%ACKNOWLEDGMENTS
We thank our colleagues on the Canadian Dark Energy Mission Study team for useful discussions and suggestions: Neil Rowlands, Ludo van Waerbeke, Justin Albert, Michael Balogh, Ray Carlberg, Pat C\^ot\'e, John Hutchings, and Dae-Sik Moon.  We acknowledge funding from the Canadian Space Agency (CSA) and the Natural Sciences and Engineering Research Council (NSERC) of Canada. High-performance computing resources for this work were supplied by the Atlantic Computational Excellence Network (ACEnet).
\vspace{5mm}

\clearpage

%% FIGURES 

\clearpage
% TABLES

\begin{table}
\begin{center}
\begin{tabular}{lccc}
\tableline\tableline
Filter & Wavelength & Obs. Time & Total Throughput \\
& (nm) & (s) & (\%) \\
\tableline 
$U$ & 300 - 440 & 542 & 0.35\\
$G$ & 440 - 550 & 542 & 0.5\\
$RIZ$ & 550 - 920 & 542 & 0.59\\
$Y$ & 920 - 1146 & 88.5 & 0.45\\
$J$ & 1146 - 1372 & 107.4 & 0.45\\
$H$ & 1372 - 2000 & 61.8 & 0.45\\
\tableline
\end{tabular}
\end{center}
\caption{\label{tab:filters}Description of filters used in simulations for Euclid. Total Throughput is estimated to include all photon losses through the system. Each filter is approximated as a box function. The $U$ and $G$ observation times listed are for the scenario where each of these filters feeds a dedicated detector; $U$ and $G$ observation times are shorter for other scenarios, as described in the text.}
\end{table}

\begin{table}
\begin{center}
\begin{tabular}{lccc}
\tableline\tableline
Filter & $m_5$(LSST) & $\gamma$ & $m_{10}$(Pan-4)\\
\tableline 
$u$ & 23.9 & 0.037 & --\\
$g$ & 25.0 & 0.038 & 25.9\\
$r$ & 24.7 & 0.039 & 25.6\\
$i$ & 24.0 & 0.039 & 25.4\\
$z$ & 23.3 & 0.040 & 23.9\\
$y$ & 22.1 & 0.040 & 22.3\\
\tableline
\end{tabular}
\end{center}
\caption{\label{tab:LSST_params}The LSST parameters used in Equation \ref{eqn:LSSTerr} (Ivezic \etal 2008) as well as the 10$\sigma$ magnitudes assumed for Pan-STARRS (Abdalla \etal 2008).}
\end{table}

%\begin{table*}
%\begin{minipage}{126mm}
\begin{table}
\begin{center}
\begin{tabular}{lcccccc}
\tableline\tableline
Scenario & Filters & Culling & $N_{gal}$ & $\sigma_{\Delta{z}/{1+z}}$ & $F_{cata}$ & FoM\\
\tableline
Euclid & $RIZ,Y,J,H$ & N & 32.2 & 0.951 & 0.3985 & $<5$\\
	    & 			 & Y & 2.3 & 0.114 & 0.0126 & $<5$\\
Euclid + Optical & $U,G,RIZ,Y,J,H$ & N & 32.2 & 0.052 & 0.0043 & 122\\
 &  & Y & 31.6 & 0.037 & 0.0016 & 126\\
Euclid + Optical (50:50 $U$:$G$ time split) & $U,G,RIZ,Y,J,H$ & N & 32.2 & 0.071 & 0.0090 & 113\\
 &  & Y & 30.8 & 0.046 & 0.0029 & 119\\
Euclid + $U$ Only & $U,RIZ,Y,J,H$ & N & 32.2 & 0.126 & 0.0414 & 85\\
 &  & Y & 25.7 & 0.062 & 0.0063 & 100\\
Euclid + $G$ Only & $G,RIZ,Y,J,H$ & N & 32.2 & 0.199 & 0.0919 & 58\\
 &  & Y & 19.7 & 0.059 & 0.0047 & 67\\
%Euclid + Optical & $U,G,Y,J,H$ & N & 32.2 & 0.118 & 0.0893 & 89\\
% &  & Y & 20.1 & 0.056 & 0.0025 & 70\\
Euclid + Pan-1 & $g,r,i,z,y,Y,J,H$ & N & 32.2 & 0.299 & 0.098 & 41\\
 &  & Y & 24.8 & 0.102 & 0.027 & 72\\
Euclid + Pan-4 & $g,r,i,z,y,Y,J,H$ & N & 32.2 & 0.175 & 0.039 & 70\\
 &  & Y & 30.1 & 0.070 & 0.010 & 105\\
Euclid + LSST & $u,g,r,i,z,y,Y,J,H$ & N & 32.2 & 0.030 & 0.0011 & 131\\
 &  & Y & 32.2 & 0.024 & 0.0003 & 133\\
Euclid + LSST with ZP errors & $u,g,r,i,z,y,Y,J,H$ & N & 32.2 & 0.032 & 0.0011 & 130\\
 &  & Y & 32.2 & 0.025 & 0.0004 & 132\\
\tableline
\end{tabular}
\end{center}
\caption{\label{tab:scenarios}Description and comparisons of various survey scenarios. Culling is defined as removing any objects with photo-$z$ error bars greater than 0.5.}
%\end{minipage}
%\end{table*}
\end{table}

\begin{table}
\begin{center}
\begin{tabular}{lc}
\tableline\tableline
Band & $\sigma_{ZP}$ \\
\tableline
$u$ & 0.05 \\
$g$ & 0.02 \\
$r$ & 0.02 \\
$i$ & 0.02 \\
$z$ & 0.03 \\
$y$ & 0.03 \\
\tableline
\end{tabular}
\end{center}
\caption{\label{tab:LSST_offset} Standard deviations of distributions from which a random zero-point offset error was chosen to apply to each pointing from LSST}
\end{table}

%% Tables may also be prepared as separate files. See the accompanying
%% sample file table.tex for an example of an external table file.
%% To include an external file in your main document, use the \input
%% command. Uncomment the line below to include table.tex in this
%% sample file. (Note that you will need to comment out the \documentclass,
%% \begin{document}, and \end{document} commands from table.tex if you want
%% to include it in this document.)

%\input tab1.tex

%% The following command ends your manuscript. LaTeX will ignore any text

\end{document}